\documentclass[preprint,12pt]{elsarticle}
\makeatletter
\def\ps@pprintTitle{%
 \let\@oddhead\@empty
 \let\@evenhead\@empty
 \def\@oddfoot{}%
 \let\@evenfoot\@oddfoot}
\makeatother
\usepackage[usenames]{color}
\usepackage[normalem]{ulem}
\usepackage{multirow,bigdelim}
\usepackage{amssymb}
\usepackage{hyperref}
\usepackage{graphicx}
\usepackage{amsmath}
\usepackage{hyperref}

\usepackage[dvips]{geometry}
\begin{document}
\begin{frontmatter}
\title{Information ratio analysis of momentum strategies}

\author[FFF]{Fernando F. Ferreira}
\ead{ferfff@usp.br}

\author[ACS]{A. Christian Silva\corref{cor1}}
\ead{csilva@idatafactory.com}

\author[JYY]{Ju-Yi Yen}
\ead{ju-yi.yen@uc.edu}


\address[FFF]{Center for Interdisciplinary Research on Complex Systems, Universidade de S\~{a}o Paulo,03828-000 S\~{a}o Paulo-SP, Brazil}
\address[ACS]{IdataFactory, Houston, Texas 77030, USA}
\address[JYY]{Department of Mathematical Sciences, University of Cincinnati, Cincinnati, Ohio 45221-0025, USA}

\begin{abstract} 
In the past 20 years, momentum or trend following strategies have become an established part of the investor toolbox. We introduce a new way of analyzing momentum strategies by looking at the information ratio (IR, average return divided by standard deviation). We calculate the theoretical IR of a momentum strategy, and show that if momentum is mainly due to the positive autocorrelation in returns, IR as a function of the portfolio formation period (look-back) is very different from momentum due to the drift (average return). The IR shows that for look-back periods of a few months, the investor is more likely to tap into autocorrelation. However, for look-back periods closer to 1 year, the investor is more likely to tap into the drift. We compare the historical data to the theoretical IR by constructing stationary periods. The empirical study finds that there are periods/regimes where the autocorrelation is more important than the drift in explaining the IR (particularly pre-1975) and others where the drift is more important (mostly after 1975). We conclude our study by applying our momentum strategy to 100 plus years of the Dow-Jones Industrial Average. We report damped oscillations on the IR for look-back periods of several years and model such oscilations as a reversal to the mean growth rate.
\end{abstract}

\end{frontmatter}

\section{Introduction}

The idea that future asset performance is a continuation of past performance is the cornerstone of momentum trading strategies. In the past 20 years, momentum has become widely accepted by both academics and practitioners as one of the strongest and most persistent factors that explain assets returns \cite{Harvey}. Even though momentum has been know since the 1930's \cite{Antonacci}, the first rigorous analysis is due to Jegadeesh and Titman \cite{JT1993}. 

Jegadeesh and Titman \cite{JT1993} construct a market neutral (long/short) portfolios of stocks based on past returns and holds the portfolio for different horizons. For instance, one takes the long position for the top decile of the best performing stocks and the short position for the bottom decile of the worst performing stocks, rebalancing the portfolio every month. One then checks if such a long/short portfolio results in significant positive returns. 

Since the work of  Jegadeesh and Titman \cite{JT1993}, momentum has been extended to different asset classes, portfolios, and other world markets. Momentum has been reported in international equity markets by \cite{Doukas,Forner,Nijman,Rouwenhorst}; in industries by \cite{Lewellen,Moskowitz}; in indexes by \cite{Bhojraj,Asness1997};  and in commodities by \cite{Erb,Moskowitz2011}.  Single risky asset momentum is analyzed in \cite{Daniel,Barberis,Hong,Berk,Johnson,Ahn,Liu,Sagi} to cite a few. Recently, momentum has also been studied linking its performance to business cycles and regimes by \cite{Chordia,Kim2013,Griffin,Guidolin,Barroso}.

On the theoretical aspect, behavioral finance is generally evoked to explain momentum \cite{Daniel,Barberis,Hong}. In general, theoretical studies suggest that if stock prices overreact or under-react to information, then the trading decision based on the past performances will occur. Daniel, Hirshleifer, and Subrahmanyam \cite{Daniel}  propose that under-reaction and overreaction are consequences of investor's overconfidence on the inside information and self-attribution bias. Barberis, Shleifer and Vishny \cite{Barberis} connect overreaction of stock prices to investor's attitude towards a series good or bad news, and under-reaction of stock prices  to investor's attitude towards information such as earning announcement. In Hong and Stein's \cite{Hong} model,  the investors are categorized into two groups, namely, the news watchers and the momentum traders, which lead to under-reaction at short horizons and overreaction at long horizons. Generally speaking, this direction of studies describe investors as Bayesian optimizers: the investor observes or receives  information at each investment time period, and updates his/her investment decision according to his/her belief. 
These behavioral models predict that under-reaction implies positive short-lag autocorrelation, and that overreaction implies negative long-lag autocorrelation. 

Alternatively to under and overreaction \cite{Daniel,Barberis,Hong} there are other causes that have been cited as possible explanation to the abnormal momentum return. Lewellen suggests that the lead-lag cross-serial correlation should explain cross-sectional momentum \cite{Lewellen}. Conrad and Kaul point to the cross-sectional variation of asset returns \cite{Conrad}. Chordia and Shivakumar and others \cite{Chordia,Kim2013,Griffin,Guidolin,Barroso} study business cycles and suggest that time-varying expected returns can explain momentum.

The traditional long/short momentum  strategy mixes portfolio construction with the technical rule (past performance, as in technical analysis \cite{Brock,Griffioen}) used to select the assets to be included in the portfolio. The portfolio construction introduces a relative performance effect since the investor needs to build a portfolio by selecting the assets based on the technical rule and assign a proportion of the total wealth to each of the asset selected. To highlight this portfolio aspect, momentum is sometimes called ``cross-sectional momentum" to contrast with "time-series momentum" introduced by Moskowitz, Ooi and Pedersen \cite{Moskowitz2011}. Time series momentum is the study of the technical rule outside a portfolio (apply momentum on individual assets): basically one of the simplest trend following strategies, and therefore a building block for more complex strategies. Moskowitz, Ooi and Pedersen perform an extensive study using over 25 years of index, futures and forward contracts' data and show that every asset they have analyzed (58 in total) present strong time series momentum.  

Similar to Moskowitz, Ooi and Pedersen, we focus on the momentum of individual assets. We study the technical rule (moving average of past returns) for one asset, therefore avoiding the portfolio effect that is important for cross-section momentum. This work adds to the paper of \cite{Moskowitz2011} by looking at the information ratio of the time series momentum strategy. Our work also contributes to the literature of linking momentum to cycles/regimes \cite{Chordia,Kim2013,Griffin,Guidolin,Barroso}. However, contrary to the previous studies, we do not associate economical episodes to the regimes. Our approach is to divide and transform the data in a way such that the final asset returns are as close as possible to stationary. We believe that our work is new in this respect.

We study momentum by looking at the risk adjusted performance measured by the information ratio (IR) as a function of the look-back lag used to construct the portfolio. Our main new contribution from a mathematical point of view, is to present in close form the risk associated with the momentum strategy. Previous works \cite{Lo1990,Lewellen,Moskowitz2011} calculate the same expression for the average return as given here, however they do not calculate the standard deviation of the strategy. Furthermore, we analyze the stability of the results across time as non-stationary effects become important in explaining the results. We find that both autocorrelation and mean drift of the random process are important in the final performance of the strategy. In particular, for look-back periods up to 4 months, the most important effect is the autocorrelation; and for look-back periods larger than 4 months to 1 year, the drift. However, in contrast with previous studies, we find that the mean drift is the most important factor after 1975.

In section \ref{sec:stationaryProcess}, we briefly introduce the notion of stationary process that is relevant for this article and describe how we find stationary regimes. In section 3, we present our momentum strategy and find the theoretical average performance, theoretical standard deviation and information ratio. Section 4 and 5 presents empirical results. We first look at stationary data comparing it to our theoretical formulas and then at non-stationary data. We confirm the general findings that momentum works for look-back (portfolio formation period) periods of up to 1 year after which the performance degrades finding a minimum for look-back periods around 2 years. Furthermore, we show that the performance improves again for look-back periods that are longer than 2 years, indicating an oscillatory (wave like) market behavior. Some formulae derivations are gathered in Appendix.

\section{Stationary Process} 
\label{sec:stationaryProcess}

The strict definition of a stationary process is that the joint probability distribution of all random variables is invariant under time shifts or translations. Equivalently, the probability density depends only on the time difference since the time origin is not relevant \cite{Feller,Gardiner}. However, it is most common to use the weak-sense stationary definition. That is, one requires that the data first moment and covariance do not change in time. In particular, the variance exists and the covariance depends only on the time difference. %Let us  point out that a nonlinear function of a strict stationary process is still strictly stationary, but this is not true for the weak stationary case.

We take the view that financial time series are not weak-sense stationary in general. This includes many of the common transformation of the price time series, such as log-returns. Most of the literature does not discuss the effect of non-stationary data; however, some studies show measurable and important effects \cite{Seemann,McCauley}.

\begin{figure}[htbp]
\centering
\includegraphics[width=\textwidth]{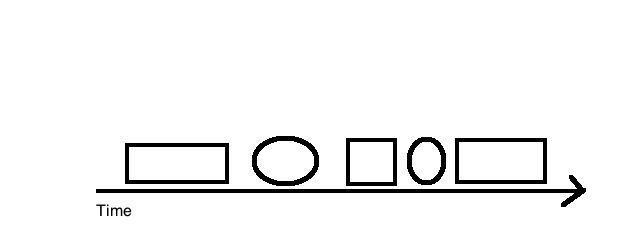}
\caption{Cartoon representation of stationary patches in a give time
series \label{fig:cartoon}}
\end{figure}

We follow \cite{Seemann} and assume that the data has patches or regimes of stationary periods. Therefore, our time series can be described by Figure \ref{fig:cartoon}. The rectangles and the circles in Figure \ref{fig:cartoon} represent intervals of the time series that are stationary. 

One example of such assumption  is the intraday FX data studied in \cite{Seemann}, another example is  the distribution of trading volumes (or number of trades) for stocks discussed in \cite{Silva}. It has been shown by \cite{Seemann,Silva} that every day the patterns repeat. As such, though the process is not stationary inside of the day (clearly since the volatility changes from parts of the day to other parts of the day), we can have a stationary process of all the first $5$ minutes (for instance) across different days.

Another possibility is that every day of the week represents a different stochastic process. That is, all Mondays are represented by a probability distribution, all Tuesdays by another, and so forth \cite{Tsiakas}. In fact, the actual probability distribution might even have a different functional form and the relation between all the days can be arbitrary. In other words, all Mondays could be correlated with each other and those could be correlated to Tuesdays and so forth.

In general, we do not postulate a periodic structure, nevertheless, we expect the presence of some repetition (Fig. \ref{fig:cartoon}) in order to have enough data to perform ensemble averages. Furthermore, in order to practically detect such periods, we shall simplify the problem by detecting periods with constant mean and constant variance. We assume that the autocorrelation is well-behaved enough to preserve the time independent property.

To detect periods with constant mean (drift), we use the ``Breaks for Additive Season and Trend" (BFAST) algorithm \cite{Verbesselt}. BFAST first starts by using the Loess regression to decompose the time series into seasonal trends and irregular components. Thereafter, the algorithm performs a loop where it uses the methodology developed by  \cite{Bai} until the number and the position of the breakpoints are unchanged. Intuitively, the algorithm  finds the trend by fitting a piecewise linearity where the breakpoints (changes from one linearity to the next) are found at the same time to the linear fit. From an implementation standpoint, we use the R package ``bfast" in our empirical studies. 

Besides BFAST, there are several other algorithms and we cite a few but we postpone the comparison for a future study. Particularly interesting is the recent work of \cite{Matteson} which is able to detect a change in the probability distribution non parametrically. There is also \cite{Fukuda} which presents a heuristic algorithm that introduces a $t$-test type of statistic. For other alternatives, see also \cite{McCauley,Seemann} and \cite{Guidolin} for models that include regime switching explicitly.
%%%%%%%%%%%%%%

\section{Model} 

The momentum strategy intends to extract predictability of future price returns from past price returns. Here we define price return by the following expression: 
\begin{equation} r_{i} = \ln (S_{i}/S_{i-1}), 
\end{equation} 

\noindent where $S_{i}$ is the price in period $i$.

In order to understand momentum strategies, we use a proxy algorithm that should represent the general characteristics of any given momentum strategy that can be implemented by buying past winners and/or taking short positions in past losers \cite{JT1993}. The momentum strategy we select is:

\begin{equation} 
\left\{ \begin{array}{lll} m_{n}(N) =
\displaystyle\sum_{i=n-N+1}^{n} r_i/N, & ~ & ~\\ \\ \mbox{if} ~ ~ m_{n}(N) > 0 &
\mbox{ Buy } & m_{n}(N) ~~\mbox{Shares} \\ \\ \mbox{if}~~ m_{n}(N) < 0 &\mbox{
Short } & m_{n}(N) ~~\mbox{Shares}
\end{array}\right. 
\label{eq:strategy} 
\end{equation} 

\noindent where $m$ is a simple moving average, and $N$ is the look-back period used to calculate the moving average. This same set of rules was used in \cite{Asness2013,Kim,Moskowitz2011,Lewellen,Lo1990} and other prior studies.\footnote{There is no guarantee that such an algorithm represents well momentum strategies, however we defer the question of how to represent a class of strategies for future study. For now, we consider that our algorithm is able to capture the main mathematical features present in a general momentum strategy.} 

In order to implement the algorithm (Eq. (\ref{eq:strategy})) described above, we first reduce the effect of volatility clustering. We do so by dividing the returns by the average absolute returns of the past $p$ periods (Eq. (\ref{eq:normalizeData})). This transformation avoids look-ahead bias and creates a strategy that can be implemented realistically.

\begin{equation}\label{eq:normalizeData} 
X_t = \frac{r_t}{\sum_{i=1}^{i=p} \mid r_{t-i}/p \mid} 
\end{equation}

The average return for our momentum strategy (Eq. (\ref{eq:strategy})) is given by

\begin{equation}\label{eq:R} 
\left \langle R \right \rangle = \frac{1}{T-N+1}
\sum_{t=N}^{T} m_{t-1}(N) X_{t} 
\end{equation}

\noindent where $N$ is the look-back period used to calculate $m_n(N)$, and $T$ is the total length of our data series (for instance the total number of weeks). By using Eq. (\ref{eq:strategy}), one can rewrite  Eq. (\ref{eq:R}) as

\begin{eqnarray} 
\left \langle R \right \rangle &=& \frac{1}{T-N+1}
\frac{1}{N}\sum_{t=N}^{T}\sum_{i=t-N}^{t-1}X_i X_t \\ \nonumber &=&
\frac{1}{T-N+1} \frac{1}{N}\left[ \sum_{t=N}^{T} X_t X_{t-N} + \sum_{t=N}^{T} X_t
X_{t-N+1} + \cdots +\sum_{t=N}^{T} X_t X_{t-1} \right]\\ \nonumber &=&\frac{1}{N}
\sum_{i=1}^{N} \left \langle X_{t} X_{t-i} \right \rangle 
\end{eqnarray}

\noindent where $<\ >$ stands for average (sometimes represented by $E[\  \ ]$). The last equality is only true if the process for $X$ is such that the product $X_T X_{T-\tau}$ is equal in probability to $X_{T-1}X_{T-1-\tau}$, and hence it depends only on $\tau$.

It was pointed in section \ref{sec:stationaryProcess} that in order to exactly model non-stationary data, we need to know how the different stationary patches relate to each other because the moving average $m(N)$ crosses different patches (Fig.\ref{fig:cartoon}). Due to this complexity, we restrict our theoretical analysis to stationary patches before working with the non-stationary data.

\subsection{Risk and Return for stationary random variables}

For stationary random variable $X_t$, the expected return (Eq. (\ref{eq:R})) can be expressed as an average of auto-covariance as follows:

\begin{equation}\label{eq:theAve} 
\left \langle R \right \rangle = \frac{1}{N}
\sum_{i=1}^{N} \left \langle X_{t} X_{t-i} \right \rangle = \mu^{2} + \frac{V}{N}
\sum_{i=1}^{N} \rho(t,t-i) 
\end{equation}

\noindent where $\rho$ is the autocorrelation function, $V$ the variance and $\mu$ the mean of the stationary stochastic process $X$. Notice that the result (Eq. (\ref{eq:theAve})) is independent of functional form of the distribution of $X$.

The variance $Var(R)$ of the strategy in Eq. (\ref{eq:strategy}) is given by:

\begin{equation} \label{eq:risk1} 
Var(R) = \left \langle \left(\frac{X_t}{N}
\sum_{i=1}^{i=N} X_{t-i}\right)^{2} \right \rangle- \left \langle
\frac{X_t}{N}\sum_{i=1}^{i=N} X_{t-i} \right \rangle^{2}. 
\end{equation}

The first term in Eq. (\ref{eq:risk1}) relates to the autocorrelation of the squared return and the cross-correlation with the squared return (similar to the leverage effect). The first term can be re-written as:

\begin{equation} \label{eq:risk2} 
\sum_{i=1}^{N} \left \langle X_t^{2}
X_{t-i}^{2}\right \rangle + \sum_{i,j=1,i \neq j}^{N} \left \langle X_t^{2}
X_{t-i}X_{t-j}\right \rangle. 
\end{equation}

We further simplify Eq. (\ref{eq:risk1}) by assuming a multivariate Gaussian distribution. Thus, the correlations are linear correlations and that the marginal distributions are Gaussian. Although empirical financial data are not described by a Gaussian distribution, the weekly normalized returns obtained according to Eq. (\ref{eq:normalizeData}) are well approximated by a Gaussian distribution.

We can therefore calculate the variance  of our momentum strategy by using the characteristic function of the multivariate Gaussian distribution. Performing the right order of differentiation and enforcing that the variance $V$ and the drift $\mu$ of the returns are constants and that the autocorrelation depends only on time lags, we have the variance of our strategy given by:

\begin{eqnarray}\label{eq:theVar} 
Var(R) &=& \frac{1}{N^{2}}\bigg[ N V^{2} + N \mu^{2} V
+ N^{2} V \mu^{2} \\ \nonumber &+&V^{2} \big(\sum_{i=1}^{N} \rho(t,t-i)\big)^{2} +V^{2}\sum_{i,j=1,i\neq
j}^{N} \rho(t-i,t-j)\\
\nonumber &+&\mu^{2} V \Big(2\sum_{i=1}^{N} \rho(t,t-i) +\sum_{i,j=1,i\neq
j}^{N}\big(\rho(t,t-j)+\rho(t-i,t-j)+\rho(t,t-i)\big)\Big) \bigg]  
\end{eqnarray}

\noindent where $\rho(t,t-i)$ is the correlation coefficient of the returns from time $t$ and $t-i$. Details of the calculation can be found in the Appendix. In the next section we will derive useful asymptotic limits of Eq. (\ref{eq:theVar}).

\subsection{Limits and interpretations}

In contrast to prior studies \cite{Kim,Lewellen,Lo1990,Moskowitz2011} we do not look only at the return of the strategy . We use the variance (Eq. (\ref{eq:theVar})) to calculate the Sharpe or Information Ratio (IR) defined here by the ratio between the average return and the standard deviation. The general expression is fairly complex but two limiting cases are enough to help us understand how the IR depends on the parameters.

\begin{figure}[htbp] \centering
\includegraphics[width=\textwidth]{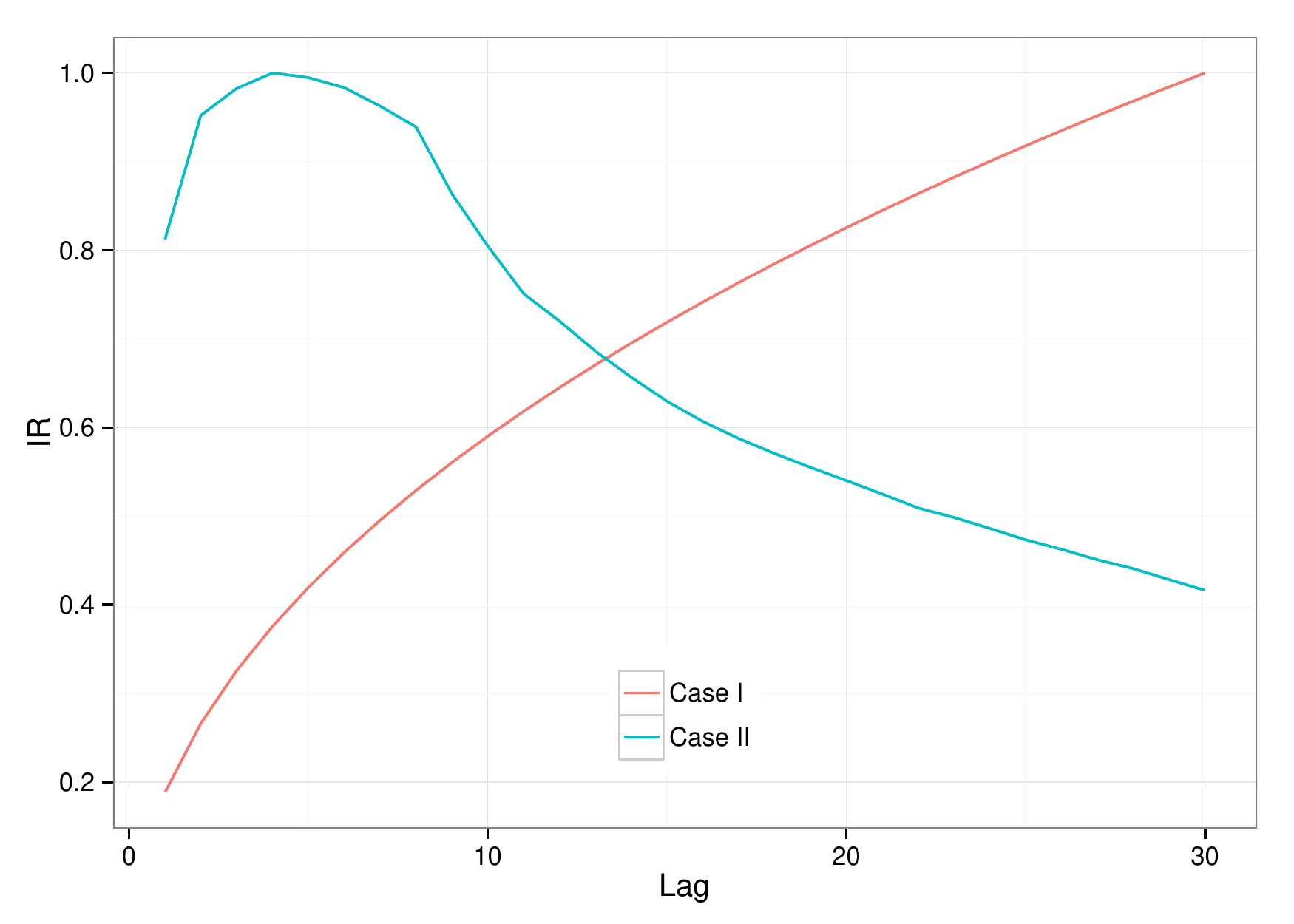} \caption{Increasing red line: IR as function of lookback lag $N$ for case I ($\rho=0$, Eq. (\ref{eq:case1})) normalized to have maximum of $1$. Decreasing blue line: Normalized IR for case II
($\mu=0$, Eq. (\ref{eq:case2})) generated using simulated data from an hypotetical process with $\rho \neq 0$ for lag 1 to lag 5. \label{fig2}} 
\end{figure}

In case I, all the autocorrelations are zero: $\rho(t,t-i)=0$. This is equivalent to say that the log-returns are independent and identically distributed (iid) Gaussian random variables. The Information Ratio (IR) is given by:

\begin{equation}\label{eq:case1} 
\text{IR} =
\frac{\mu^{2}}{\sqrt{V\mu^{2}+\frac{V^{2}}{N}+\frac{\mu^{2}V}{N}}} 
\ \xrightarrow[N \to \infty]{} \
\frac{\left|\mu\right|}{\sqrt{V}}, 
\end{equation} 

\noindent where we also show the behavior when $N \rightarrow \infty$. $N \rightarrow \infty$ is the limit of long (or short) and hold since $m(N)$ converges to $\mu$. It is interesting that the optimal point is when $N \rightarrow \infty$; in this case, the best information ratio is in fact what one would expect for a given process: mean over standard deviation. Any other $N$ gives worst results. Hence, as expected, if  the given process is iid, the moving average $m(N)$ provides one way to estimate $\mu$. A cartoon representation of the IR as a function of $N$ for case I is given in Fig. \ref{fig2}.

In case II, we assume that $\mu=0$. Thus, all performance comes from autocorrelation. The IR is given by

\begin{equation}\label{eq:case2} \text{IR} =
\frac{\sum_{i=1}^{N}\rho(t,t-i)}{\sqrt{N+(\sum_{i=1}^{N}\rho(t,t-i))^{2}+(\sum_{i,j=1,i\neq
j}^{N}\rho(t-j,t-i))}} \end{equation}

\noindent where the exact shape of IR as a function of $N$ depends on the way $\rho$ is a function of $N$. Practically it is very unlikely that $\sum_{i=1}^{N} \rho(t,t-i)$ grows fast enough to dominate the $\sqrt{N}$ term in the denominator. More surprising is that Eq. (\ref{eq:case2}) does not depend on the variance $V$, in other words, the IR of the strategy is the same for very large or small $V$. The expression is also useful practically since one can calculate IR for a given correlation value. For instance, for $N=2$, $\rho(t,t-1) = 0.05$ and $\rho(t,t-2) = 0.02$, IR $ \approx 0.0422$ per period, that means that if we are dealing with weekly data, for instance, IR $=\sqrt{52}*0.0422= 0.3$ per year.

In order to illustrate case II, we look at a moving average process ($MA$ i.e. of the kind $x(t) = a_{1} \epsilon_{1}(t)+a_{2} \epsilon_{2}(t-1)+..., \epsilon_{i}=N(0,\sigma)$) with autocorrelation $\rho \neq 0$ from lag 1 to lag 5. The general shape of case II is presented in Figure \ref{fig2} (case II, blue line). We have chosen to show an arbitrary $MA$ process because it is the only process from the $ARMA$ family where the $\rho$ term in the numerator can grow fast enough before $1/\sqrt{N}$ in the denominator takes over. This creates the ``hump'' in the graph. The hump is located where the $MA$ process has autocorrelation different from zero and the size of the hump depends on the autocorrelation intensity.

In summary, case I (red line) is increasing with the value of $N$, while  case II (blue line)  increases initially and then decreases slowly (Figure \ref{fig2}).

It is clear that if we have a pure case II, there could be an optimal $N$ above which we get worst risk adjusted performance. Therefore, it is generally not advantageous to use a momentum type strategy unless the best $N$ is selected. 

The dependence of the IR on $N$ is different for case I and case II: the first case grows and the second decreases in the limit of large $N$.
Normally, empirical data has IR as a hybrid of case I and case II, that indicates that a pure momentum strategy will transit from a case II dominated performance to a case I dominated performance with the increase of $N$. Large $N$ shows that case I is targeted: momentum estimates the drift. Keep in mind  that if the sum of the autocorrelations happens to be positive (negative), case I is shifted up (down) due to the case II contribution.

Finally, we remind the readers that our discussion here assumes a stationary process. That is clearly not the case in practice. We assume that there are patches or periods where the data is approximately stationary. Next, we will look at data by approximately finding these stationary patches. We will conclude by looking at the full non-stationary data set and present a simple model that describes momentum for non-stationary data.

%%%%%%%%%%%%

\section{Empirical stationary analysis}
\label{sec:E_stationary} 

We use the Dow-Jones Industrial Average (DJIA) Index from 05/1896 to 02/2013 to perform our empirical studies.\footnote{We present the study with the DJIA index but we also looked at the S\&P 500 index from 1950 to the present with virtually the same results, therefore we will not show them here.} The daily DJIA index values are downloaded from Federal Reserve Economic Data (FRED) webpage.\footnote{Data download uses "quantmod" library in R \cite{R}} We convert the daily index values to weekly index values and calculate weekly log-returns. We choose to work with weekly data because we get at least 4 times more data than the traditional monthly values and we also reduce considerably any market micro-structure issue from daily data.

Most prior studies have a minimum holding period of one month even when using weekly data \cite{Kim}. In contrast with most prior studies,  we  re-balance our position weekly (holding period of one week), which maximizes data usage. All our analyses are done using the  software {\bf R} \cite{R}.  A simple sample code pertinent to this article can be found in \cite{Rpubs}.

We take the view that financial log-returns are generally not stationary \cite{Guidolin,McCauley,Seemann}. Notice that this assumption is different from several prior studies, take for instance \cite{Lehmann,Lo1988,Lo1990} where the results are obtained with 25 years of data (1962 to 1986). This study is more in line with studies that advocate business cycles dependence such as \cite{Chordia}.

%%%%%%%%%%%%

\begin{figure}[htbp] \centering
\includegraphics[width=\textwidth]{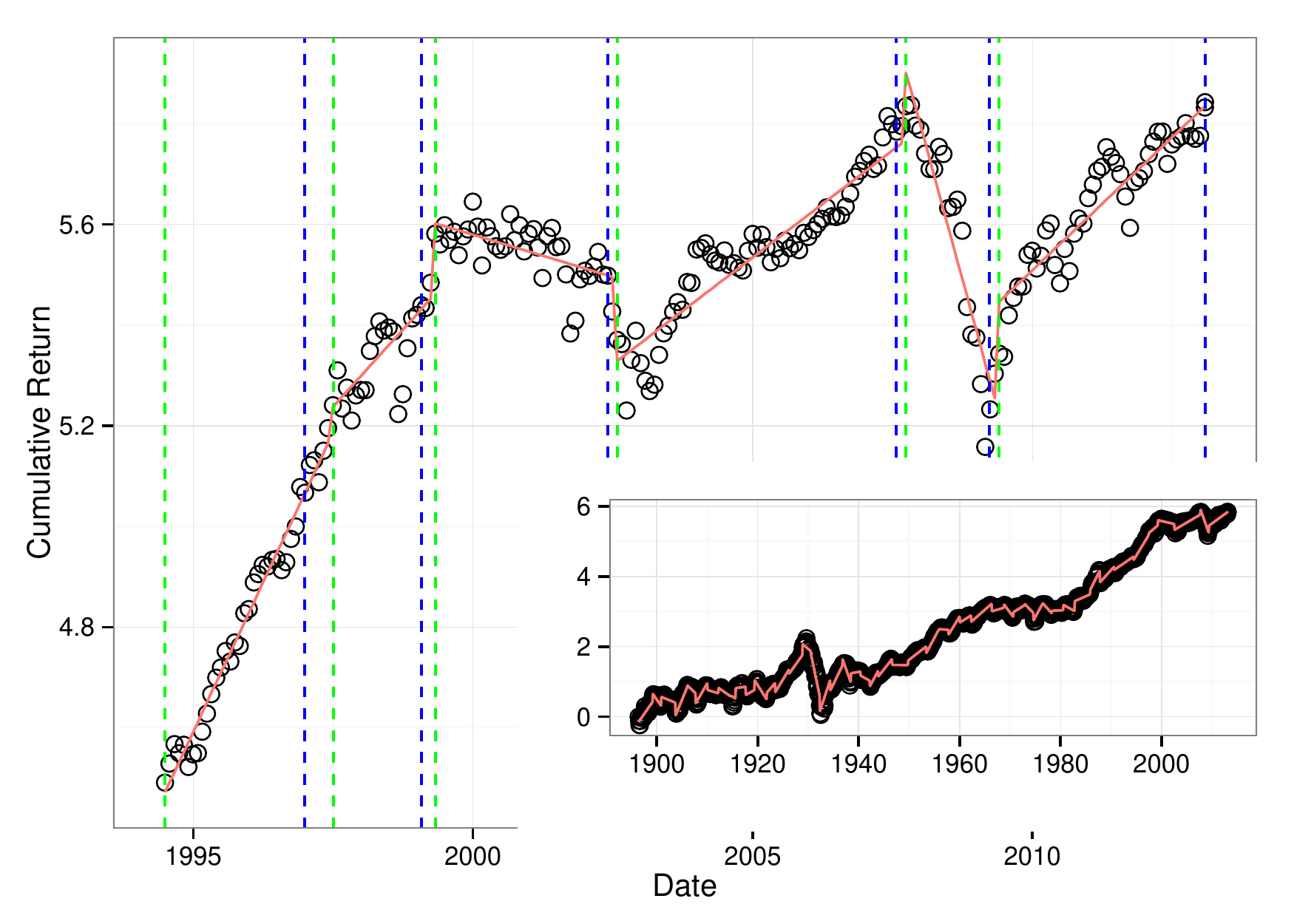} \caption{Log cumulative
return of weekly DJIA from 1995 to 2012 (points). The intervals between green and red vertical dash lines were obtained by
using the BFAST algorithm which finds these intervals by fiting piecewise linear functions to the data. The linear fits are represented by the solid red lines. \label{fig4}} \end{figure}

%%%%%%%%%%%%

In order to find stationary periods, we first transform the data using Eq.
(\ref{eq:normalizeData}) to obtain approximately Gaussian data with constant volatility, and then we search for periods of constant drift. We use the ``Breaks for Additive Season and Trend''" (BFAST) algorithm to find the constant drift periods. Here we are interested in the trend component which is a piecewise linear function where the breakpoints (change from a linear function to the next) are determined by BFAST as well. 

We apply BFAST to the log DJIA index sampled monthly. We choose a monthly index because running the algorithm is faster and because we want periods that are few years long in order to have enough data points (weeks) within each regime. Our average regime length is of 2.2 years with a maximum of approximately 5 years, and a minimum of nearly 1 year.

After applying BFAST to find the stationary patches, we ignore any regime with less than 1.3 years (70 weeks). We are left with 47 regimes (patches).  We plot the log cumulative return for DJIA from 1995 to 2013 together with the fitted trends in Figure \ref{fig4}. Each regime is defined between dashed green line and the consecutive dashed blue line. The inset shows the data from 1900 to 2013.

In general, linear regression in Figure \ref{fig4} shows a very good fit. Some patches present a good adherence with the liner fit while others present a larger fluctuation around the linear fit. The goodness of the fit validates the method used in this work. Patches with larger fluctuations correspond to troubled economical moments as in 1997, 2002 and 2008.

\begin{figure}[htbp] \centering
\includegraphics[width=\textwidth]{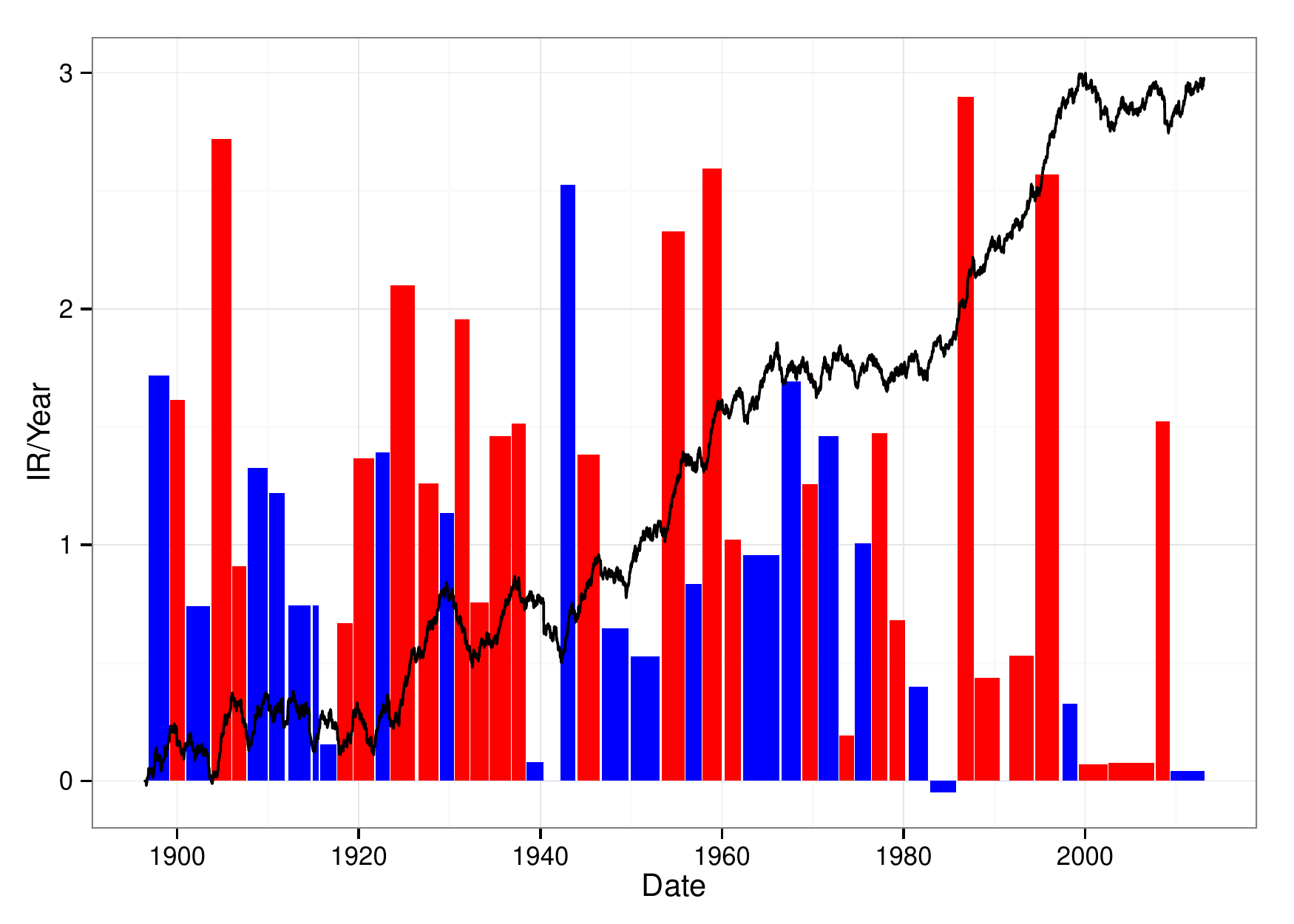} \caption{Time series representation of stationary periods. Bar graph shows the annualized IR values for each patch/regime.
The red bars refer to case I and blue bars refer case II. We also show the log of the rescalled DJIA index (black line) in arbitrary units. \label{fig5}}
\end{figure}

We calculate the empirical performance of the strategy (Eq. (\ref{eq:strategy})) for look-back periods ranging from $N=1$ to $N=43$ weeks within each regime. Figure \ref{fig5} shows the rescaled log-DJIA plotted together with the maximum annualized (weekly values multiplied by $\sqrt{52}$)  information ratio. The color coded bar graph refers to case I and case II in Figure \ref{fig2}. If the maximum IR is for look-back periods $N>20$ weeks we classify it to be a case I (red) otherwise a case II (blue) period. That is, the red bars imply a graph IR versus lag $N$ that is visually closer to case I than case II (Figure \ref{fig2}). In addition to Figure \ref{fig5},  Table \ref{table1} includes the standard error for the IR within each regime.  

From Figure \ref{fig5}, we notice that  case I is more prominent for post 1975 data. This is due to two reasons. First,  the DJIA increases quite steadily from 1980 to 2000 which implies a large drift effect. Second, after 1975, the autocorrelation effect in Eq. (\ref{eq:theAve}) is mostly negative ($\sum_{i} \rho(t,t-i) < 0$) or at best insignificant. Consequently, there is a higher IR with  large $N$ (better to buy and hold - case I) compared to  small $N$ (case II - cumulative autocorrelation). 

\begin{figure}[ht] 
\centering \includegraphics[height=8cm, width=\textwidth]{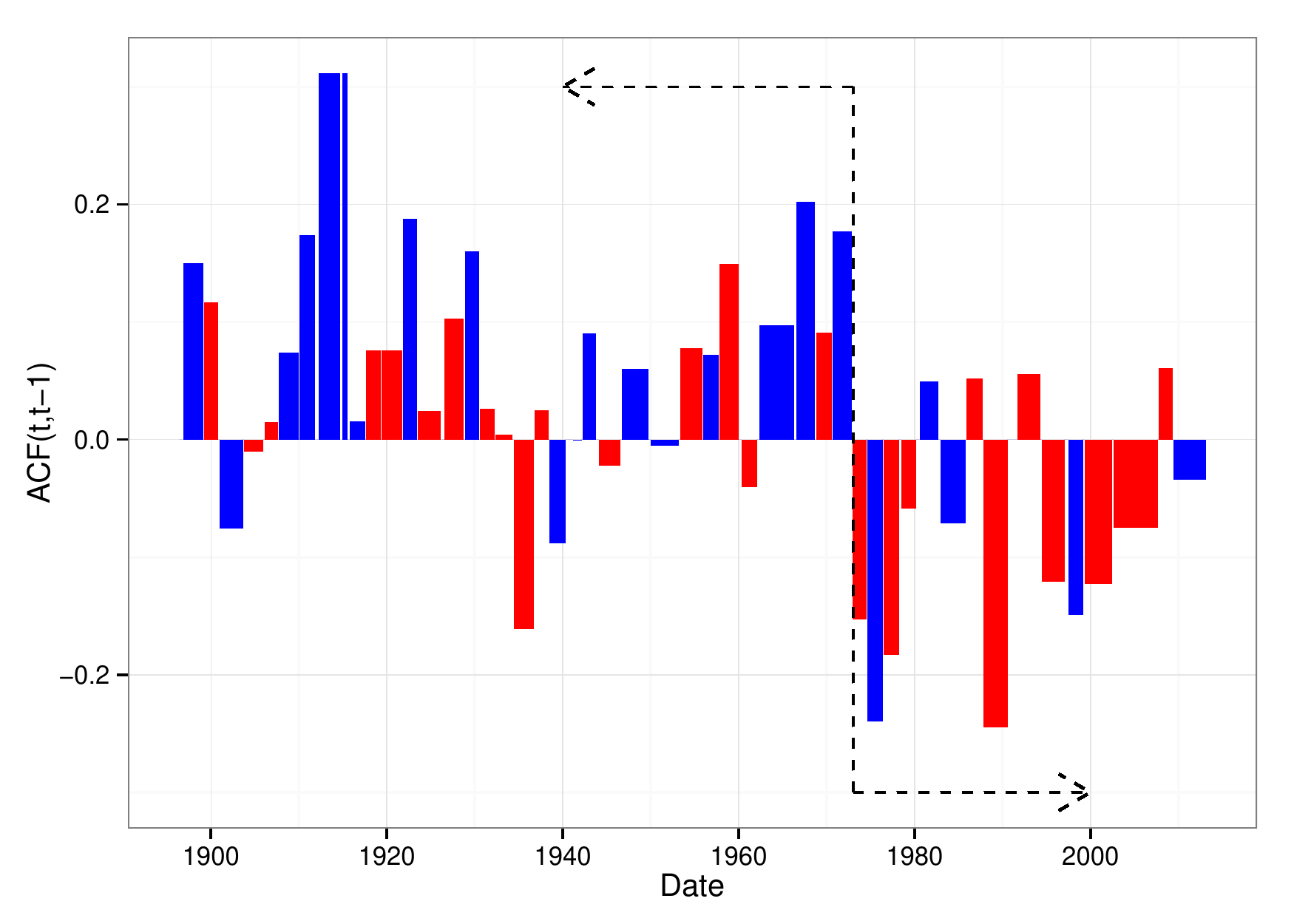}
\caption{First lag autocorrelation for rescaled DJIA log-returns within each regime. Red bars correspond to case I and blue to case II using the same convention that was used in Figure \ref{fig5} \label{fig6}} 
\end{figure}

Figure \ref{fig6} and Table \ref{table1} present the first lag autocorrelation within each regime. The bars are again color coded to match Figures \ref{fig2} and \ref{fig5}. Notice that from 1900 to 1975, most bars are positive, and after 1975 the bars are mostly negative. The sign of the first lag autocorrelation survives even if we calculate the autocorrelation over all 4080 weeks up to 1975 and all 1988 weeks after 1975. The autocorrelation for all weekly returns before 1975 is 0.17 and after 1975 it becomes $-0.04$. Therefore, even though most of the autocorrelation values per regime are marginally significant (Table \ref{table1}) there is a likely change on the sign of the autocorrelation of weekly returns in 1975.

Priviously, work by Lo and MacKinelay  \cite{Lo1988,Lo1990} hint to the autocorreltion regime change we reported here (Figure \ref{fig6}). They show that for an index of large cap stocks, the autocorrelation for the sub-period (1975$-$1985) is not significantly positive, where as, it is significantly positive from 1962 to 1988. See also the work of Froot and Perold \cite{Froot}, which study the first lag autocorrelation for returns from 15 minutes to 1 week for stock indices with data up to 1990. In agreement with our results, they show that the large positive autocorrelation decreases substantially after 1970 going negative (see Figure 5 in \cite{Froot}).

To further compare our results to Lo and MacKinelay \cite{Lo1988,Lo1990}, we calculate the first four lag autocorrelations of the weekly log-returns from July 6, 1962 to December 31, 1987 for the DJIA index. Our values have the same sign as reported by  Lo and MacKinelay \cite{Lo1990} but different magnitudes. We get 1.1\%, 1.7\%, 5.1\%, $-$2.8\% from lags 1 to 4 whereas they have 7.4\%, 0.7\%, 2.1\% and $-$0.5\% (for the value-weighted CRSP index, see Table 1 of \cite{Lo1990}). We attribute these different values to the fact that, we use the DJIA and they look at the value-weighted Center for research in security prices (CRSP) index.

Furthermore, the autocorrelation values shown in Figure \ref{fig6} and Table \ref{table1} are calculated with the DJIA rescaled log-returns (Eq. (\ref{eq:normalizeData})). Therefore we expect an additional difference especially on the magnitude of the autocorrelation. We find that the first four autocorrelation of our rescaled DJIA index are  5.7\%,1.6\%,0.7\% and -3.2\%  which agree in sign with both Lo and MacKinelay (7.4\%, 0.7\%, 2.1\% and $-$0.5\% ) and with the un-scaled DJIA log-returns (1.1\%, 1.7\%, 5.1\%, $-$2.8\%) .  The positive sign of the first autocorrelation only shows that by using data from (1962$-$1988) we are missing the regime change around 1975. In other words, by taking the data from 1962 to 1988, we are mixing our 11 regimes in such a way that the first lag autocorrelation becomes positive. That is, the pre-1975 data ``dominates" the post-1975 data when the autocorrelation of the entire period (1962 to 1988) is calculated.

For completeness, it is worth mentioning that \cite{Bondt1989,Lehmann} report that individual weekly stock returns are negatively autocorrelated using data from virtually the same period (1962 to 1986) as in \cite{Lo1990}. This apparent dilemma (since indices are made of individual stocks) was addressed by \cite{Lo1990}. They show that it is possible for the autocorrelation of the index to be positive even though the autocorrelation of the component stocks is negative if the stocks cross-autocorrelation (lead-lag relation among stocks) is large and positive. 

Our autocorrelation results in Figure \ref{fig6} indicate a fundamental change for at least large cap stocks that start in 1975. Not only is the first lag autocorrelation of DJIA weekly returns negative but  assuming that first lag autocorrelation for individual stocks is still negative (see \cite{Gutierrez} with data from 1983 to 2003), cross-autocorrelation between stocks should be negative or insignificant after 1975.

%%%%%%%%%%%

%%%%%%%%%%%%
%%%%%%%%%%

\begin{figure}[htbp] \centering
\includegraphics[width=\textwidth]{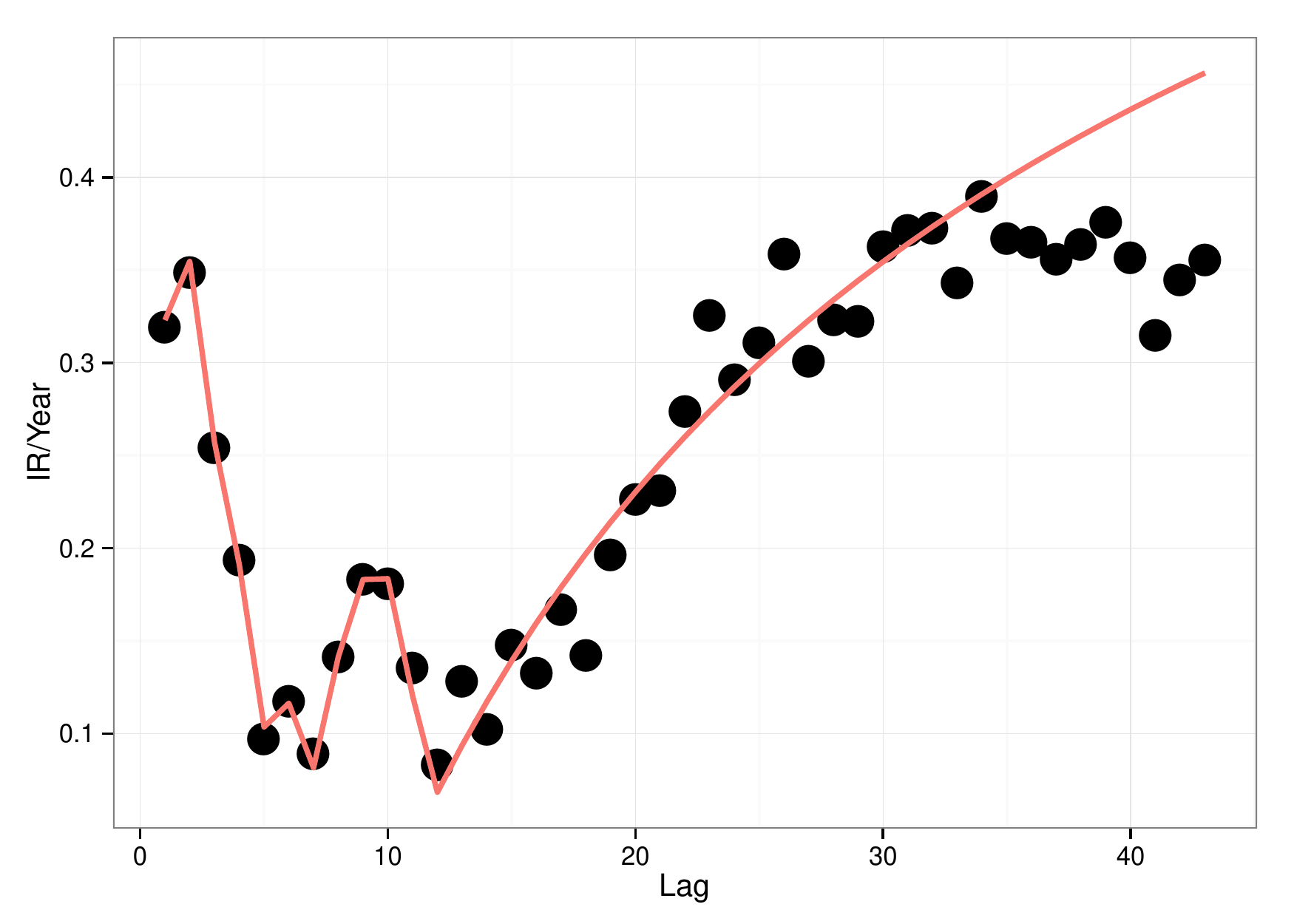} \caption{Average IR vs. look-back lag
over the 47 patches determined by BFAST algorithm. Circles represent the data and
the red line is the best fit theoretical IR given by Eqs. (\ref{eq:theAve}) and
(\ref{eq:theVar}).
\label{fig:aveSpec}} 
\end{figure}

%%%%%%%%%%%%%%%%%%%%%%%
\clearpage
\begin{table}
\centering
\small
\begin{tabular}{rrrrrr}
  \hline
 Regimes& \# Weeks & Acf(1) & Acf SE(95\%)  & Max IR (Year) & IR SE(95\%)  \\ 
  \hline
1896-11-06 to 1899-01-27 & 117.00 & 0.15 & 0.18 & 1.72 & 1.33 \\ 
  1899-03-30 to 1900-09-28 & 79.00 & 0.12 & 0.23 & 1.61 & 1.62 \\ 
  1901-01-04 to 1903-07-31 & 135.00 & -0.08 & 0.17 & 0.74 & 1.24 \\ 
  1903-10-02 to 1905-11-24 & 113.00 & -0.01 & 0.19 & 2.72 & 1.36 \\ 
  1906-02-02 to 1907-07-26 & 78.00 & 0.01 & 0.23 & 0.91 & 1.63 \\ 
  1907-10-04 to 1909-11-26 & 113.00 & 0.07 & 0.19 & 1.33 & 1.36 \\ 
  1910-02-04 to 1911-09-29 & 87.00 & 0.17 & 0.21 & 1.22 & 1.55 \\ 
  1912-03-29 to 1915-06-25 & 151.00 & 0.31 & 0.16 & 0.74 & 1.17 \\ 
  1915-10-01 to 1917-06-29 & 92.00 & 0.02 & 0.21 & 0.16 & 1.50 \\ 
  1917-08-31 to 1919-03-28 & 83.00 & 0.08 & 0.22 & 0.67 & 1.58 \\ 
  1919-05-29 to 1921-08-26 & 118.00 & 0.08 & 0.18 & 1.37 & 1.33 \\ 
  1921-11-04 to 1923-04-27 & 78.00 & 0.19 & 0.23 & 1.39 & 1.63 \\ 
  1923-06-29 to 1926-01-29 & 136.00 & 0.02 & 0.17 & 2.10 & 1.24 \\ 
  1926-07-30 to 1928-08-31 & 110.00 & 0.10 & 0.19 & 1.26 & 1.38 \\ 
  1928-11-30 to 1930-05-29 & 79.00 & 0.16 & 0.23 & 1.14 & 1.62 \\ 
  1930-08-01 to 1932-02-26 & 83.00 & 0.03 & 0.22 & 1.96 & 1.58 \\ 
  1932-04-29 to 1934-03-29 & 100.00 & 0.00 & 0.20 & 0.76 & 1.44 \\ 
  1934-06-01 to 1936-08-28 & 118.00 & -0.16 & 0.18 & 1.46 & 1.33 \\ 
  1936-10-30 to 1938-04-29 & 79.00 & 0.03 & 0.23 & 1.51 & 1.62 \\ 
  1938-07-01 to 1940-03-29 & 92.00 & -0.09 & 0.21 & 0.08 & 1.50 \\ 
  1942-04-02 to 1943-09-24 & 78.00 & 0.09 & 0.23 & 2.52 & 1.63 \\ 
  1944-02-04 to 1946-06-28 & 126.00 & -0.02 & 0.18 & 1.38 & 1.28 \\ 
  1946-10-04 to 1949-08-26 & 152.00 & 0.06 & 0.16 & 0.65 & 1.17 \\ 
  1949-12-30 to 1953-01-30 & 162.00 & -0.01 & 0.16 & 0.53 & 1.13 \\ 
  1953-05-29 to 1955-10-28 & 127.00 & 0.08 & 0.18 & 2.33 & 1.28 \\ 
  1955-12-30 to 1957-08-30 & 88.00 & 0.07 & 0.21 & 0.84 & 1.54 \\ 
  1957-11-01 to 1959-11-27 & 109.00 & 0.15 & 0.19 & 2.60 & 1.38 \\ 
  1960-04-29 to 1961-12-29 & 88.00 & -0.04 & 0.21 & 1.02 & 1.54 \\ 
  1962-05-04 to 1966-03-25 & 204.00 & 0.10 & 0.14 & 0.96 & 1.01 \\ 
  1966-07-29 to 1968-07-26 & 105.00 & 0.20 & 0.20 & 1.69 & 1.41 \\ 
  1968-11-01 to 1970-06-26 & 86.00 & 0.09 & 0.22 & 1.26 & 1.56 \\ 
  1970-09-04 to 1972-09-29 & 109.00 & 0.18 & 0.19 & 1.46 & 1.38 \\ 
  1972-12-01 to 1974-06-28 & 83.00 & -0.15 & 0.22 & 0.19 & 1.58 \\ 
  1974-08-30 to 1976-04-30 & 88.00 & -0.24 & 0.21 & 1.01 & 1.54 \\ 
  1976-07-02 to 1978-02-24 & 87.00 & -0.18 & 0.21 & 1.48 & 1.55 \\ 
  1978-06-30 to 1980-01-25 & 83.00 & -0.06 & 0.22 & 0.68 & 1.58 \\ 
  1980-08-01 to 1982-08-27 & 109.00 & 0.05 & 0.19 & 0.40 & 1.38 \\ 
  1982-12-03 to 1985-09-26 & 148.00 & -0.07 & 0.16 & -0.05 & 1.19 \\ 
  1985-11-29 to 1987-08-28 & 92.00 & 0.05 & 0.21 & 2.90 & 1.50 \\ 
  1987-10-30 to 1990-06-29 & 140.00 & -0.24 & 0.17 & 0.44 & 1.22 \\ 
  1991-08-30 to 1994-03-31 & 136.00 & 0.06 & 0.17 & 0.53 & 1.24 \\ 
  1994-07-01 to 1996-12-27 & 131.00 & -0.12 & 0.17 & 2.57 & 1.26 \\ 
  1997-07-03 to 1999-01-29 & 83.00 & -0.15 & 0.22 & 0.33 & 1.58 \\ 
  1999-04-30 to 2002-05-31 & 162.00 & -0.12 & 0.16 & 0.07 & 1.13 \\ 
  2002-08-02 to 2007-07-27 & 261.00 & -0.07 & 0.12 & 0.08 & 0.89 \\ 
  2007-09-28 to 2009-03-27 & 79.00 & 0.06 & 0.23 & 1.52 & 1.62 \\ 
  2009-05-29 to 2013-02-01 & 193.00 & -0.03 & 0.14 & 0.04 & 1.04 \\ 
   \hline
\end{tabular}
\caption{First lag autocorrelation and IR per regime.} 
\label{table1}
\end{table}
\clearpage
%%%%%%%%%%%%%%%%%%%%%%%%%%

Results from the autocorrelation (Figure \ref{fig6}) are further sustained by Figure \ref{fig:aveSpec}. Figure \ref{fig:aveSpec} shows the average of IR vs $N$ (look-back) over all 47 regimes. In addition to the ensemble average, we show the theoretical IR that is constructed by dividing the theoretical average performance  (Eq. (\ref{eq:theAve})) by the theoretical standard deviation of the performance (Eq. (\ref{eq:theVar})). 

The parameters for theoretical model are found by least square fitting the empirical data points. The model uses the empirical standard deviation ($1.5$)  of our renormalized data and 11 fitting parameters: the drift $\mu$ and the first 10 autocorrelations. Clearly, we are over-fitting since the number of parameters is large giving that we have only 43 data points; however, our goal here is not to find a robust model but to qualitatively show the potential of the theoretical model. 

The average IR shows that for very short look-back, 1 or 2 weeks, we have a large positive autocorrelation effect (case II). The decrease from the peak in two weeks is much faster then the natural case II would predict (Figure \ref{fig2}). Based on our fit, what explains the fast decrease from week 3 to 7 is a progression of negative autocorrelation which pull the average IR down from its maximum in two weeks. The influence of autocorrelation can be detected all the way to week 12 with a quick alternation of what looks like a ``peak" of positive autocorrelation at approximate 10 weeks ($\approx$2.5 months) in the mist of negative autocorrelation. For look-back periods of more than 16 weeks ($\approx$4 months) the IR curves clearly resembles case I, where if we add more look-back weeks we do better.

It is very unlikely that the high autocorrelation effect for the first 12 weeks is significant in the recent history, considering that most first lag autocorrelations are negative (at best neutral) after 1975 (Figure \ref{fig6} and Table \ref{table1}). However, it is true that on average both case I and case II (Figure \ref{fig2}) can be present and equally important in an average stationary regime. What is interesting is that even though both appear to be equally significant, there is a clear transition from one (case II for small $N$) to the other (case I for large $N$) as we change $N$. Due to such transition, one can classify a momentum strategy based on the portfolio formation period (look-back $N$). The 3 to 4 months look-back leads to momentum of the kind associated to under-reaction since there seams to be significantly positive autocorrelation \cite{JT2001} before 4 months. On the other side, momentum for look-back periods larger than 4 months is mostly due to the natural drift present in the asset. In the literature this drift is sometimes associated with macroeconomic variables and business cycles  \cite{Chordia,Kim2013,Griffin,Guidolin,Barroso,Conrad} and often perceived as orthogonal to behavioral models \cite{Chordia,Conrad}.

Finally, we note that for large lag $N$, the average IR in Figure \ref{fig:aveSpec} disagrees with the theoretical fit. It is difficult to draw conclusions since we stop at 10 months and the number of data points become small. In order to capture large $N$ effects (which show overreaction \cite{Bondt1989}) and also in order to be closer to the way a momentum strategy is typically tested in applications, we will construct IR versus lag $N$ for all 6068 weeks of data in the next section.

\section{Empirical non-stationary analysis}
\label{sec:NE_stationary}
In this section we apply the momentum strategy of Eq. (\ref{eq:strategy}) to all DJIA data. We do not try to break the data into stationary periods, but we still normalize the weekly log-returns using Eq. (\ref{eq:normalizeData}). The goal here is to present the effect of non-stationary data to the performance of momentum strategies. Since we normalize the log-returns using Eq. (\ref{eq:normalizeData}), we still expect the data to have constant variance. Nevertheless, the data will not have constant drift nor constant autocorrelation. Therefore, we do not expect this section to conform with case I or case II in Figure \ref{fig2}.

\begin{figure}[ht] 
\begin{minipage}[b]{0.45\linewidth}
 \centering \includegraphics[height=10cm, width=\textwidth]{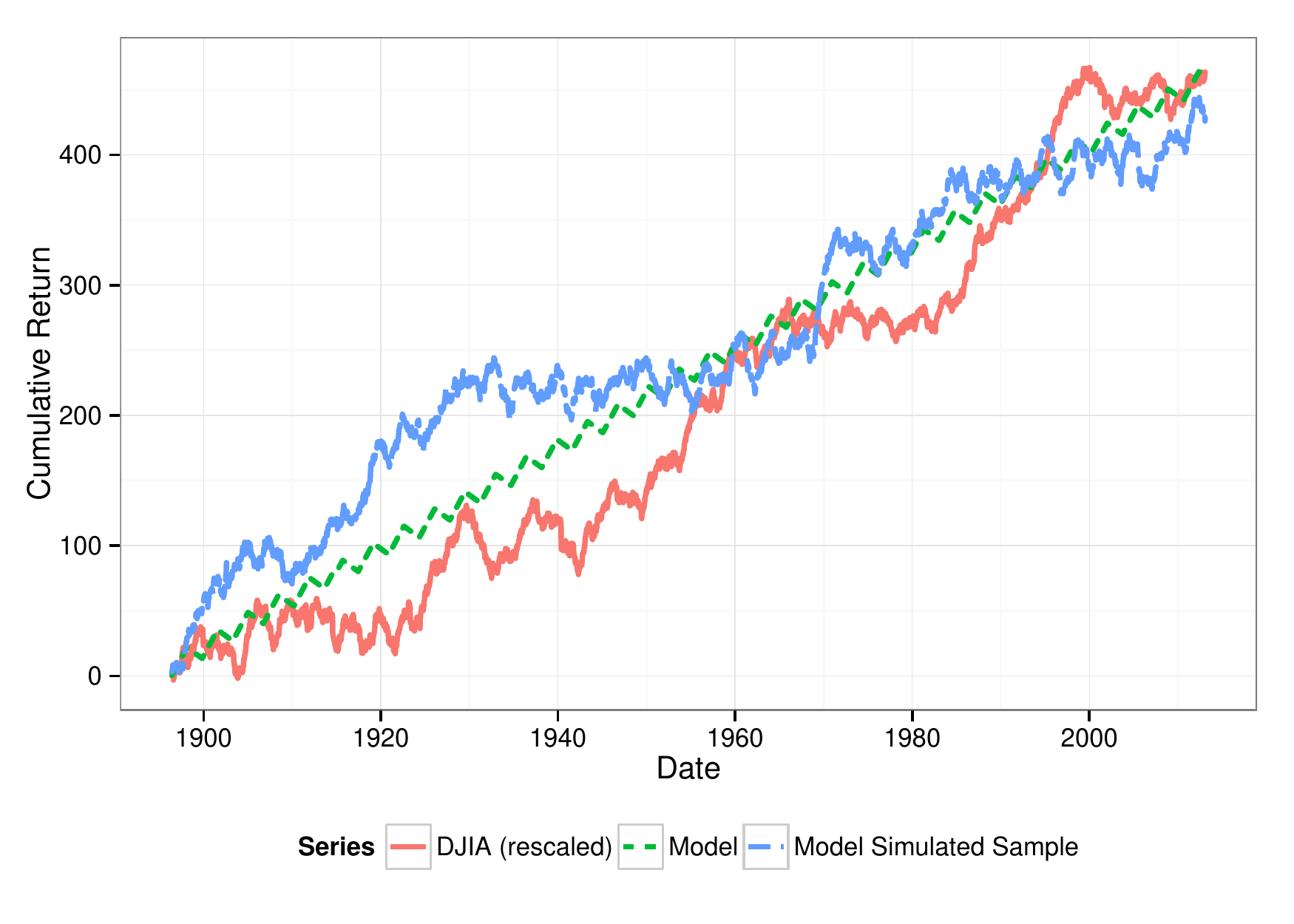}
  \caption{Rescaled DJIA index, model and Monte Carlo simulations. Solid red line shows the rescaled log-DJIA data. The dashed green line shows the model of Eq. (\ref{eq:nonSmodel}) without noise (integrated square wave). The blue line shows one Monte Carlo realization of Eq. (\ref{eq:nonSmodel}).} \label{fig:notStat2} \end{minipage} \hspace{0.1cm} \begin{minipage}[b]{0.45\linewidth} \centering
\includegraphics[height=11cm, width=\textwidth]{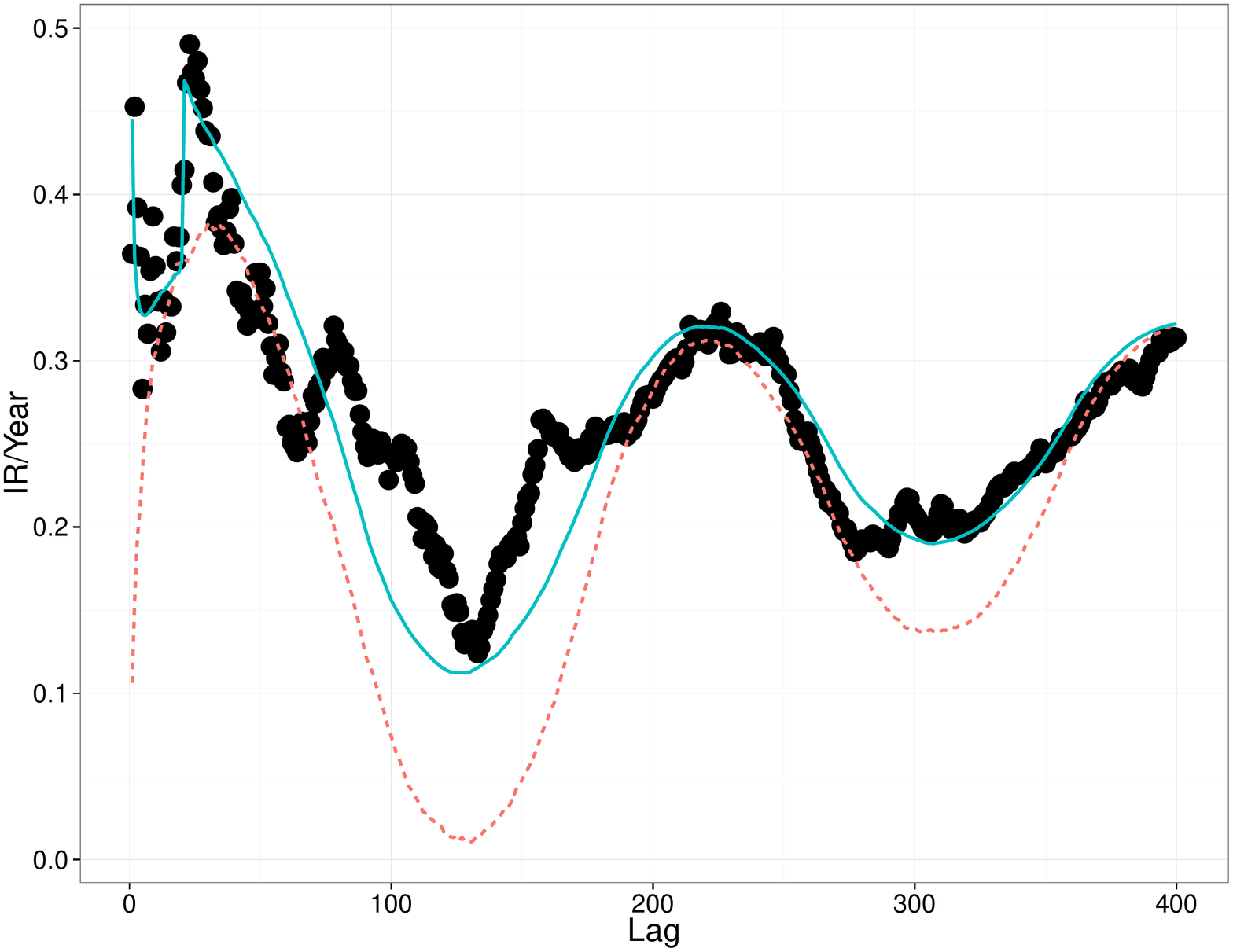} \caption{IR versus lag for 100 years of DJIA index. Dashed red line shows the IR calculated via Monte Carlo method for model in Eq. (\ref{eq:nonSmodel}) with $\epsilon$ drawn from IID gaussian distribution. Solid blue line corresponds to theoretical line from Eq. (\ref{eq:nonSmodel}) with $\epsilon$ from a MA process with autocorrelation at lag 1 and lag 20. Black circles corresponds to the DJIA rescaled data.} \label{fig:notStat} 
\end{minipage} 
\end{figure}

Figure \ref{fig:notStat} shows the information ratio (IR) versus $N$ number of weeks used to construct the moving average (Eq. (\ref{eq:strategy})). Since we have 6068 weeks (data points) we present the IR for our momentum strategy that looks back up to 400 weeks (almost 8 years). In agreement with our expectations, the curve of the IR dependence on the look-back period is not even qualitatively similar to case I or case II. However, we can still use results of the previous section as guide to develop a model here.

We learned that both autocorrelation and drift are important to explain the IR of a momentum strategy (Figure \ref{fig:aveSpec}). We know that case I is more important than case II for large time lags $N$ (portfolio formation look-back lags) since the IR in case I increases with lag and in case II it decreases (Figure \ref{fig2}). Considering this, we postulate that the oscillations are mostly due to changes in the drift of case I. That leads us to assume that the theoretical model for the normalized log-returns is given by:

\begin{equation}
\label{eq:nonSmodel}
 r_m(t) = \mu + A\:{\rm sgn}( \sin(2 \pi t/T))+\epsilon,
\end{equation}

\noindent where $\mu$ stands for the average growth rate, $A$ for the amplitude of our square wave, $T$ for the period and $\epsilon$ for the noise. Figure \ref{fig:notStat2} shows the cumulative return of our model (with and without noise added) together with the DJIA data.

We take the parameter $\mu$ to be equal to the empirical growth rate of the weekly re-normalized DJIA ($\mu=0.075$ per week). The parameter $A$ and $T$ are selected to fit the empirical IR oscillation in Figure \ref{fig:notStat}. We find that $T$ is approximately equal to 3.5 years (180 weeks) and that $A\approx 2\times\mu$. The root mean square $\sigma$ of the noise $\epsilon$ is equal to the empirical noise level of the re-normalized DJIA weekly returns over the entire period ($\sigma = 1.5/week$). 

The model in Eq. (\ref{eq:nonSmodel}) shows the long term reversals reported by \cite{Bondt1985,Bondt1987,Bondt1989,Fama} for stocks and indices. The period of 3.5 years implies a mean reversion of about 1.75 years which is within the prior reported value of 1.5 to 5 years \cite{Bondt1989}. In terms of correlations, our model has a small positive autocorrelation that progressively goes negative and back up again: oscillating with the period $T$ even when $\epsilon$ is independent and identically distributed noise . However since the empirical noise level is between 7 to 20 times larger than $\mu \pm A$ such correlations are not easy to detect by calculating the autocorrelation function. Consider Figure \ref{fig:notStat2} Monte Carlo generated sample curve for our model (blue line): notice that it is impossible to recognize the periodic oscillations. 

In order to compute the IR versus the lag length $N$ for our theoretical model (Eq. (\ref{eq:nonSmodel})), we perform Monte Carlo (MC) simulations (500 realizations). We consider two types of noise $\epsilon$. The first one is an iid Gaussian random variable with zero drift and variance equal to the empirical variance of the re-normalized data. This is the simplest case and assumes that all the performance for a momentum strategy is only due to the drift ($\mu$). The second one is a MA process with zero mean and variance as the Gaussian iid random variable, but with best fit positive autocorrelation for lag 1 (0.05) and lag 20 (0.08). The second model takes in consideration that short lags might be strongly influenced by autocorrelation (case II situation) and therefore both case I and case II co-exist as in Figure \ref{fig:aveSpec}. 

The MA process is selected by first choosing the non-zero autocorrelation lags and then by fitting the model to the empirical data (IR vs N curve) to find the numerical coefficients, which imply a correlation of 0.05 and 0.08 for lag 1 and lag 20.  We select the non-zero lags by using the intuition developed for the stationary model and by keeping the number of non-zero lags minimal. From Eq. (\ref{eq:case2}) and Figure \ref{fig2}, we know that a MA model generates a localized hump in the graph IR versus lag $N$ in the case of stationary data. The location of the hump is centered at the location of the autocorrelation. Assuming that this approximately valid here, the IR vs N  for uncorrelated $\epsilon$ can be shifted up in the most relevant places to get a better fit. Therefore, by looking at the empirical IR vs N of Figure \ref{fig:notStat}, we postulate that the most relevant lags should be around lag 1 and lag 20, since we need to shift the theoretical IR vs N generated with $\epsilon$ iid Gaussian (dashed red) at least at lag 1 and lag 20, to produce the IR vs N from the MA generated $\epsilon$ (solid blue).

Figure \ref{fig:notStat} shows the empirical IR versus lag $N$ (circles) and 2 theoretical lines (dashed red and solid blue). The best fit is achieved with $\epsilon$ drawn form a $MA$ process (solid blue line). The worst fit is with $\epsilon$ drawn from an iid Gaussian noise (red dashed line). The oscillations are well captured  by the uncorrelated $\epsilon$, especially for very large lags $N$, however it underestimates the IR ratio for small lags. We have shown (Fig. \ref{fig6})  that before 1975 the autocorrelation is significantly positive, and since we are in effect averaging the data by treating the 100 years of the DJIA as stationary, we need to include large positive short range autocorrelation to account for data before 1975. The result is the much better fit of the solid blue line in Figure \ref{fig:notStat}.

Based on Figure \ref{fig:aveSpec} one could expect to find a small autocorrelation effect on the IR for lag $N \approx 20$, which is not what we find by fitting the IR here (Figure \ref{fig:notStat}). However, this is misleading, since we would be comparing approximately stationary data to non-stationary data. Furthermore, even if some regimes have "large" lag 20 autocorrelation, this autocorrelation would have been averaged out. That is, the difference between Figure \ref{fig:aveSpec} and Figure \ref{fig:notStat}, is that Figure \ref{fig:aveSpec} is constructed by taking an average over all the IR versus lag $N$ curves created within our stationary patches. In other words, all IR for $N  \approx 20$ that have a hump are averaged with all IR which do not show a hump. Since at $N \approx 20$ the contribution to IR of the autocorrelation hump is of order $1/\sqrt{20}$ and the contribution of drift is of order $1$ (Eq. (\ref{eq:theVar})), the average is dominated by the drift. The result is that, the effect of the autocorrelation for lags larger than 15 in Fig. \ref{fig:aveSpec} should not be evident. 

The model of Eq. (\ref{eq:nonSmodel}) with $\epsilon$ drawn from a MA process not only fits the empirical IR curve well (Figures \ref{fig:notStat} and \ref{fig:notStat2}), but it also agrees well with prior studies in the literature. It shows a large positive first lag autocorrelation of the normalized weekly returns which agree with \cite{Lo1988,Lo1990}. It also shows a large autocorrelation around 1/2 year, which suggests that autocorrelation is the main driver of traditional momentum returns for look-back of 3 to 12 months and that the drift plays as a much less important role \cite{JT2001,JT2002}. Our finding here however suggests, that the results in the literature are mostly due to data prior to 1975.

\section{Conclusions}

In the present work we perform statistical analysis on a momentum strategy and find a closed form formula for the expected value and the variance. We are therefore able to present an analytic expression for the Information Ratio. The innovation here allows one to compute and discuss the risk adjusted performance of a momentum strategy.

The theory developed here assumes that the time series are patches stationary. One key point is to find the proper stationary patch. Once this condition is satisfied, we get good agreement between theory and data as discussed in section \ref{sec:E_stationary}. Notwithstanding, when a longer time series is considered, the stationary assumption may no longer be valid. In this scenario, we proposed a stochastic model in section \ref{sec:NE_stationary}. This model agrees well with the data by introducing a periodic change in the average growth rate of the DJIA. 

Summarizing our empirical results: in section \ref{sec:E_stationary}, we find that both autocorrelation and drift can be important in describing the risk adjusted performance of a time series momentum strategy. More originally, we show that the information ratio appears to present 2 phases. The first phase is for short look-back periods and the second for long look-back periods. The first phase is mainly driven by autocorrelation where as the second phase by average return (drift). Furthermore, there is a behavioral change at 1975, were the first lag autocorrelation of weekly returns go from mainly positive before 1975 to mainly negative after 1975.

In section \ref{sec:NE_stationary}, we find an oscillatory IR for long portfolio formation periods which we associate with periodic cycles in the market. However, we emphasize that the results in section \ref{sec:NE_stationary} are inherently unstable since we use 100 years of DJIA data. That means that if we repeat the same study for different sub-samples, we will find different quantitative and sometimes qualitative results. It is true that for 100 years ($\approx 6000$ weeks) of data our results are statistically significant with little doubt. In fact the exact extreme case is the study done in section \ref{sec:E_stationary}, where we have $\approx 100$ weeks per patch and therefore the error bars are  $\approx 10$ times larger. This is one of the challenges of empirical finance: to keep significance we may end up mixing stationary patches that creates an average behavior which may hide the true mechanism. On the other hand if we look at less data we may have to compromise significance which makes us more reluctant to believe in the observed effects. Balancing these two is a great challenge when dealing with non-stationary data.

\newpage

\section{Appendix}

The characteristic function for the multivariate Gaussian random variable is

\begin{equation}
\phi(s) =  \exp\Big( i\sum_{i} s_{i} \mu_{i} -\frac{1}{2} \sum_{i} V_{ii} s_{i}^{2} + \frac{1}{2} \sum_{i \neq j} s_{i} s_{j} V_{ij}\Big )
\end{equation}
We will calculate each term of Eq. (\ref{eq:risk2}) separately. The summand of the first term is
\begin{equation}
\label{eq:riskA1}
\left \langle X_t^{2} X_{t-i}^{2}\right \rangle =\left \langle X_{k}^{2} X_{i}^{2}\right \rangle = \frac{\partial^{2}}{\partial s_{k}^{2}} \frac{\partial^{2}}{\partial s_{i}^{2}} \phi(s) |_{s=0}
\end{equation}
where we have simplified our notations by taking current time $t$ as $k$; and
past times being labeled as $i$, $j$ or $q$. The double partial derivatives in (\ref{eq:riskA1}) result into:

\begin{eqnarray}
\label{eq:cal1}
V_{ii} V_{kk} \phi(s) \\ \nonumber
 -V_{ii} [i \mu_{k}-V_{kk} s_{k}-\sum_{j} s_{j}V_{ij}]^{2} \phi(s) \\ \nonumber
+V_{ik}^{2} \phi(s) \\ \nonumber
-V_{ik}
 [i\mu_{k}-V_{kk}s_{k}-\sum_{j}S_{j}V_{kj}][i\mu_{i}-V_{ii}s_{i}-\sum_{j} s_{j}V_{ij}]\phi(s) \\ \nonumber
+V_{ik}^{2} \phi(s) \\ \nonumber
-V_{ik}
 [i\mu_{k}-V_{kk}s_{k}-\sum_{j} s_{j} V_{kj}][i\mu_{i}-V_{ii}s_{i}-\sum_{j} s_{j}V_{ij}]\phi(s) \\ \nonumber
-V_{ik}
 [i\mu_{k}-V_{kk}s_{k}-\sum_{j} s_{j} V_{kj}][i\mu_{i}-V_{ii}s_{i}-\sum_{j} s_{j}V_{ij}]\phi(s) \\ \nonumber
-V_{ik}
 [i\mu_{k}-V_{kk}s_{k}-\sum_{j} s_{j} V_{kj}][i\mu_{i}-V_{ii}s_{i}-\sum_{j} s_{j}V_{ij}]\phi(s) \\ \nonumber
-V_{kk}
[i\mu_{i}-V_{ii} s_{i}-\sum_{j} s_{j} V_{ij}]^{2} \phi(s) \\ \nonumber
+[i \mu_{k} - V_{kk} s_{k} - \sum_{j} s_{j} V_{kj}]^{2} [i \mu_{i} - V_{ii} s_{i} - \sum_{j} s_{j} V_{ij}]^{2} \phi(s) \nonumber
\end{eqnarray}

\noindent where $j$ is the index of  the summation. Let $s=0$, thus
Eq. (\ref{eq:cal1}) is reduced to:

\begin{equation}
\left \langle X_{k}^{2} X_{i}^{2}\right \rangle=V_{ii}V_{kk}+2V_{ik}^{2}+V_{ii}\mu_{k}^{2}+4V_{ik}\mu_{i}\mu_{k}
+\mu_{i}^{2}V_{kk}+\mu_{i}^{2}\mu_k^{2}
\end{equation}

\noindent Now, the summand of the second term of Eq. (\ref{eq:risk2}) is given by

\begin{equation}
\label{eq:riskA2}
\left \langle X_t^2X_{t-i}X_{t-j}\right \rangle = \left \langle X_{k}^{2} X_{i}X_{j}\right \rangle = \\
\frac{\partial^{2}}{\partial s_{k}^{2}} \frac{\partial}{\partial s_{i}} \frac{\partial}{\partial s_{j}} \phi(s) |_{s=0}
\end{equation}
Notice that we use $j$ to indicate a different time lag in Eq. (\ref{eq:riskA2}). The partial derivatives in Eq. (\ref{eq:riskA2}) result into:

\begin{eqnarray}
\label{eq:cal2}
+V_{ij}V_{kk}\phi(s)\\ \nonumber
-V_{ij} [i \mu_{k} - V_{kk} s_{k} - \sum_{q} s_{q} V_{kq}]^{2} \phi(s) \\ \nonumber
+ V_{ik}V_{kj} \phi(s) \\ \nonumber
-V_{ik} [i \mu_{k} - V_{kk} s_{k} - \sum_{q} s_{q} V_{kq}] [i \mu_{j} - V_{jj} s_{j} - \sum_{q} s_{q} V_{jq}] \phi(s) \\ \nonumber
+V_{ik}V_{kj} \phi(s)\\ \nonumber
-V_{jk} [i \mu_{k} - V_{kk} s_{k} - \sum_{q} s_{q} V_{kq}] [i \mu_{i} - V_{ii} s_{i} - \sum_{q} s_{q}V_{iq}] \phi(s) \\ \nonumber
-V_{jk} [i \mu_{k} - V_{kk} s_{k} - \sum_{q} s_{q} V_{kq}] [i \mu_{i} - V_{ii} s_{i} - \sum_{q} s_{q}V_{iq}] \phi(s) \\ \nonumber
-V_{ik} [i \mu_{k} - V_{kk} s_{k} - \sum_{q} s_{q} V_{kq}] [i \mu_{j} - V_{jj} s_{j} - \sum_{q} s_{q} V_{jq}] \phi(s) \\ \nonumber
-V_{kk} [i \mu_{i} - V_{ii} s_{i} - \sum_{q} s_{q} V_{iq}] [i \mu_{j} - V_{jj} s_{j} - \sum_{q} s_{q} V_{jq}] \phi(s) \\ \nonumber
+ [i \mu_{k} - V_{kk} s_{k} - \sum_{q} s_{q} V_{kq}]^{2} [i \mu_{j} - V_{jj} s_{j} - \sum_{q} s_{q} V_{jq}] \\ \nonumber
\times [i \mu_{i} - V_{ii} s_{i} - \sum_{q} s_{q} V_{iq}]  \phi(s) \\ \nonumber
\end{eqnarray}

\noindent Let $s=0$, then (\ref{eq:cal2}) becomes

\begin{equation}
V_{ij}V_{kk}+2V_{ik}V_{jk}+2V_{kj}\mu_{i}\mu_{k}
+V_{ij}\mu_{k}^{2}+2V_{ik}\mu_{j}\mu_{k}
+V_{kk}\mu_{j}\mu_{i}+\mu_{k}^{2} \mu_{i} \mu_{j}
\end{equation}

\noindent Hence, the first term of (\ref{eq:risk1}), i.e. the square of strategy return (R), is given by

\begin{eqnarray}
\label{eq:cal3}
\left \langle R^{2} \right \rangle &=& \frac{1}{N^2}\Big(
\sum_{i} \big(V_{ii}V_{kk}+2V_{ik}^{2}+V_{ii}\mu_{k}^{2}+4V_{ik}\mu_{i}\mu_{k}
+\mu_{i}^{2}V_{kk}+\mu_{i}^{2}\mu_k^{2}\big)\nonumber \\ 
&+&\sum_{i,j, i \neq j} \big(V_{ij}V_{kk}+2V_{ik}V_{jk}+2V_{kj}\mu_{i}\mu_{k}
+V_{ij}\mu_{k}^{2}+2V_{ik}\mu_{j}\mu_{k}+  %\nonumber
V_{kk}\mu_{j}\mu_{i}+\mu_{k}^{2} \mu_{i} \mu_{j}\big)\Big) %\nonumber
\end{eqnarray}

\noindent Now,  we compute the second term of Eq. (\ref{eq:risk1}), i.e. the square of the average, and we get

\begin{eqnarray}
\label{eq:cal4}
\left \langle R \right \rangle^{2} &=& \frac{1}{N^2}\Big(\sum_{i} \big(V_{ij} +\mu_{i}\mu_{j}\big)\Big)^{2} \nonumber \\&=&
\frac{1}{N^2}\Big(\sum_{i} \big(V_{ik}^{2} + 2V_{ki}\mu_{k}\mu_{i}+\mu_{k}^{2}\mu_{i}^{2}\big)  \nonumber \\
&+&\sum_{i,q,i\neq q} \big(V_{ki}V_{kq}+V_{ki}\mu_{k}\mu_{q}+V_{kq}\mu_{k}\mu_{q}
+\mu_{k}^{2}\mu_{i}\mu_{q}\big)\Big)
\end{eqnarray}
\noindent where we use the same simplifying notation ($i$,$j$,$k$) to
represent the time lags.

Finally, from  Equations (\ref{eq:cal3}) and (\ref{eq:cal4}), we obtain
Eq. (\ref{eq:risk1}) as the following:

\begin{eqnarray}
\label{eq:finalGeneralVar}
Var(R) &=& \frac{1}{N^2}\Big(\sum_{i} \big(V_{ii}V_{kk}+V_{ik}^{2}+V_{ii}\mu_{k}^{2}+2V_{ki}\mu_{i}\mu_{k}+V_{kk}\mu_{i}^{2}\big) \\ \nonumber
&+&\sum_{i,j,i\neq j} \big(V_{ij}V_{kk}+V_{ki}V_{kj}+V_{kj}\mu_{i}\mu_{k}+V_{ij}\mu_{k}^{2}+V_{kk}\mu_{i}\mu_{j} +V_{ik}\mu_j\mu_k\big)\Big) \nonumber
\end{eqnarray}
Now, assuming  that the variance $V$ and the drift $\mu$ are constants and that the autocorrelation depends only on time lag, (\ref{eq:finalGeneralVar}) can be converted to Eq. \ref{eq:theVar} .

\begin{eqnarray}
\label{eq:theVarApp} Var(R) &=& \frac{1}{N^{2}}\bigg[ N V^{2} + N \mu^{2} V
+ N^{2} V \mu^{2} \\ \nonumber &+&V^{2} \big(\sum_{i=1}^{N} \rho(t,t-i)\big)^{2} +V^{2}\sum_{i,j=1,i\neq
j}^{N} \rho(t-i,t-j)\\
\nonumber &+&\mu^{2} V \Big(2\sum_{i=1}^{N} \rho(t,t-i) +\sum_{i,j=1,i\neq
j}^{N}\big(\rho(t,t-j)+\rho(t-i,t-j)+\rho(t,t-i)\big)\Big) \bigg]  
\end{eqnarray}

\noindent where we convert back to the time lag notation used in the main text by reverting the $i,j,k$ notation. For example, we take $V_{ki} = V \rho(t,t-i)$,$V_{ij} = V \rho(t-i,t-j)$ and $V_{kk}=V$.

\section*{Acknowledgments}

ACS would like to thank the participants of R/Finance 2013 for comments and the organizers for financial support to present a preliminary version of this paper.
FFF acknowledges financial support from Funda\c{c}\~ao Amparo a Pesquisa do Estado de S\~ao Paulo (FAPESP) and Funda\c{c}\~ao Instituto de F\'isica Te\'orica (FIFT) for hospitality. We thank Constantin Unanian for detailed comments.

\end{document}